\documentclass[12pt]{revtex4}

\usepackage[english]{babel}
\usepackage{graphicx,rotating}
\usepackage{fleqn,a4p,here}
\usepackage{lep2rep}

\newcommand{\Qfbhad}{\ensuremath{{Q^{\mathrm{had}}_{\mathrm{FB}} }}}

\def\gappeq{\mathrel{\rlap {\raise.5ex\hbox{$>$}} {\lower.5ex\hbox{$\sim$}}}}
\def\lappeq{\mathrel{\rlap{\raise.5ex\hbox{$<$}} {\lower.5ex\hbox{$\sim$}}}}

\begin{document}
\pagestyle{empty}
\begin{flushright} 
%{\tt hep-ph/0406270}\\
{CERN-PH-TH/2004-116}\\
\end{flushright}
\vspace*{5mm}
\begin{center} 
{\large\bf The Electroweak Interactions in the Standard Model and beyond}\\
\vspace*{0.8cm} {\bf Guido Altarelli}\\
\vspace{0.6cm} {}
CERN, Department of Physics, Theory Division\\
 CH--1211 Geneva 23, Switzerland \\
\vspace*{0.9cm} 
{\bf Abstract}
\end{center}
\noindent 
We present a concise review of the status of the Standard Model and of the models of new physics. 
 \vspace{1.9cm}
\vspace*{0.5cm}
\begin{center}
{\it Talk given at the ICFA School on\\
Instrumentation in Elementary Particle Physics\\
Itacuruca, Rio de Janeiro, Brazil, December 2003}
\end{center}
\vfill
\noindent
\begin{flushleft} 
CERN-PH-TH/2004-116\\ 
June 2004 \end{flushleft}
 \eject 
 \setcounter{page}{1} 
 \pagestyle{plain}
 %\noindent 2/4
 \section{Precision Tests of the Standard Model} 
 The results of the electroweak precision tests as well as of the searches
for the Higgs boson and for new particles performed at LEP and SLC are now available in nearly final form.  Taken together
with the measurements of $m_t$, $m_W$ and the searches for new physics at the Tevatron, and with some other data from low
energy experiments, they form a very stringent set of precise constraints \cite{LEPEW} to compare with the Standard Model (SM) or with
any of its conceivable extensions. When confronted with these results, on the whole the SM performs rather
well, so that it is fair to say that no clear indication for new physics emerges from the data \cite{AG}. 

%..............................................................................
\begin{table}[h]
\begin{center}
  \renewcommand{\arraystretch}{1.30}
\begin{tabular}{|ll||r||r|}
\hline
&Observable& {Measurement}  & {SM fit}  \\
\hline
\hline
&$\MZ$ [\GeV{}] & $91.1875\pm0.0021\pz$ &91.1873$\pz$ \\
&$\GZ$ [\GeV{}] & $2.4952 \pm0.0023\pz$ & 2.4965$\pz$ \\
&$\shad$ [nb]   & $41.540 \pm0.037\pzz$ &41.481$\pzz$ \\
&$\Rl$          & $20.767 \pm0.025\pzz$ &20.739$\pzz$ \\
&$\Afbzl$       & $0.0171 \pm0.0010\pz$ & 0.0164$\pz$ \\
\hline
&$\cAl$~(SLD)   & $0.1513 \pm0.0021\pz$ & 0.1480$\pz$ \\
\hline
&$\cAl~(P_\tau)$& $0.1465 \pm0.0033\pz$ & 0.1480$\pz$ \\
\hline
&$\Rbz{}$       & $0.21644\pm0.00065$   & 0.21566     \\
&$\Rcz{}$       & $0.1718\pm0.0031\pz$  & 0.1723$\pz$ \\
&$\Afbzb{}$     & $0.0995\pm0.0017\pz$  & 0.1037$\pz$ \\
&$\Afbzc{}$     & $0.0713\pm0.0036\pz$  & 0.0742$\pz$ \\
&$\cAb$         & $0.922\pm 0.020\pzz$  & 0.935$\pzz$ \\
&$\cAc$         & $0.670\pm 0.026\pzz$  & 0.668$\pzz$ \\
\hline
&$\swsqeffl$
  ($\Qfbhad$)   & $0.2324\pm0.0012\pz$  & 0.23140     \\
\hline
\hline
&$\MW$ [\GeV{}]
                & $80.425\pm0.034\pzz$  &80.398$\pzz$ \\
&$\GW$ [\GeV{}]
                & $ 2.133\pm0.069\pzz$  & 2.094$\pzz$ \\
\hline
&$\Mt$ [\GeV{}] ($\pp$~\cite{TEVEWWG-top})
                & $178.0\pm4.3\pzz\pzz$ &178.1$\pzz\pzz$ \\
\hline
& $\dalhad$~\cite{bib-BP01}
                & $0.02761\pm0.00036$   & 0.02768     \\
\hline
%-----------------------------------------------------------------------
\end{tabular}\end{center}
\caption[Overview of results]{ Summary of electroweak precision
measurements at high $Q^2$~\cite{LEPEWWG:2003}. The first block shows
the Z-pole measurements.  The second block shows additional results
from other experiments: the mass and the width of the W boson measured
at the Tevatron and at LEP-2, the mass of the top quark measured at
the Tevatron, and the contribution to $\alqed$ of the hadronic
vacuum polarisation. }
\label{tab:msm:input}
\end{table}
%..............................................................................

All electroweak Z pole measurements, combining the results of the 5
experiments, are summarised in Table~\ref{tab:msm:input}. Information on the Z partial widths are contained in the quantities:
\begin{eqnarray}
\shad & = & \frac{12\pi}{\MZ^2}\ \frac{\Gee\Ghad}{\GZ^2}\,, \qquad
\Rl ~ = ~ \frac{\shad}{\slept} ~ = ~ \frac{\Ghad}{\Gll}\,, \qquad
\Rq ~ = ~ \frac{\Gqq}{\Ghad} \,.
\end{eqnarray}
Here $\Gll$ is the partial decay width for a pair of massless charged
leptons.  The partial decay width for a given fermion species are related to the effective vector and axial-vector coupling
constants of the neutral weak current:
\begin{eqnarray}
\Gff & = & N_C^f \frac{\GF\MZ^3}{6\sqrt{2}\pi} 
\left( \gaf^2 C_{\mathrm{Af}} + \gvf^2 C_{\mathrm{Vf}} \right) 
          +  \Delta_{\rm ew/QCD}\,,
\end{eqnarray}
where $N_C^f$ is the QCD colour factor, $C_{\mathrm{\{A,V\}f}}$ are
final-state QCD/QED correction factors also absorbing imaginary
contributions to the effective coupling constants, $\gaf$ and $\gvf$
are the real parts of the effective couplings, and $\Delta$ contains
non-factorisable mixed corrections.

Besides total cross sections, various types of asymmetries have been
measured.  The results of all asymmetry measurements are quoted in
terms of the asymmetry parameter $\cAf$, defined in terms of the real
parts of the effective coupling constants, $\gvf$ and $\gaf$, as:
\begin{eqnarray}
\cAf & = & 2\frac{\gvf\gaf}{\gvf^2+\gaf^2} ~ = ~ 
           2\frac{\gvf/\gaf}{1+(\gvf/\gaf)^2}\,, \qquad
\Afbzf ~ = ~ \frac{3}{4}\cAe\cAf\,.
\end{eqnarray}
The measurements are: the forward-backward asymmetry ($\Afbzf =
(3/4)\cAe\cAf$), the tau polarisation ($\cAt$) and its forward
backward asymmetry ($\cAe$) measured at LEP, as well as the left-right
and left-right forward-backward asymmetry measured at SLC ($\cAe$ and
$\cAf$, respectively).  Hence the set of partial width and asymmetry
results allows the extraction of the effective coupling constants. 
In particular, from the measurements at the Z, lepton universality of the neutral weak current was
established at the per-mille level.

Using the effective electroweak mixing angle, $\swsqefff$, and the
$\rho$ parameter, the effective coupling constants are given by:
\begin{eqnarray}
\gaf              & = & \sqrt{\rho}~T^f_3\,, 
\qquad 
\frac{\gvf}{\gaf} ~ = ~ 1-4|q_f|\swsqefff\,,
\end{eqnarray}
where $T^f_3$ is the third component of the weak iso-spin and $q_f$
the electric charge of the fermion. The effective electroweak mixing
angle is thus given independently of the $\rho$ parameter by the ratio
$\gvf/\gaf$ and hence in a one-to-one relation by each asymmetry
result.

\begin{figure}[htb]
\begin{center}
\includegraphics[width=0.8\textwidth]{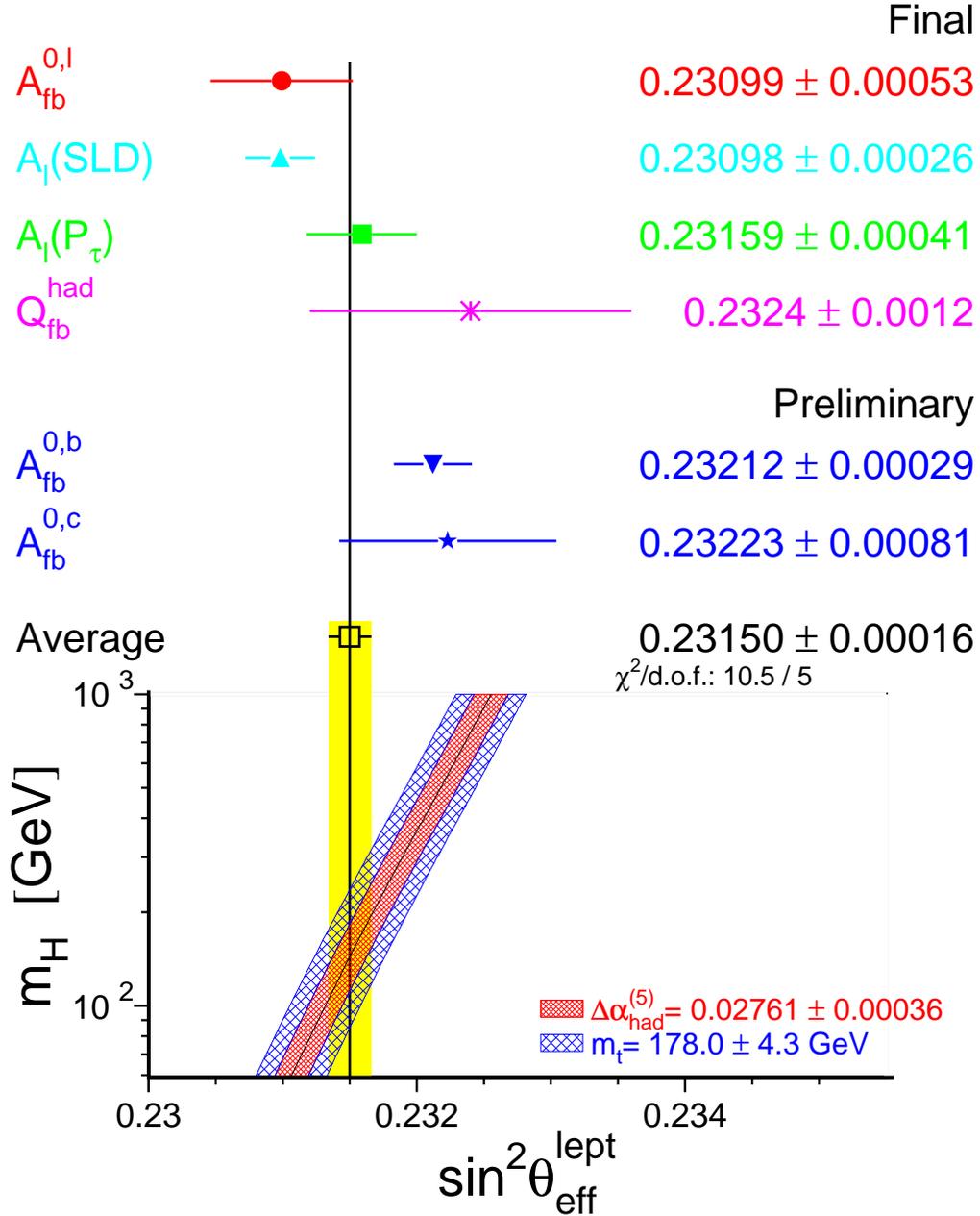}
\caption[Effective electroweak mixing angle] { Effective electroweak
mixing angle $\swsqeffl$ derived from measurement results depending on
lepton couplings only (top) and also quark couplings (bottom) \cite{LEPEW}.  Also
shown is the prediction of $\swsqeffl$ in the SM as a function of
$\MH$, including its parametric uncertainty dominated by the
uncertainties in $\dalhad$ and $\Mt$, shown as the bands. }
\label{fig:sin2teff} 
\end{center}
\end{figure}

The various asymmetries determine the effective electroweak mixing
angle for leptons with highest sensitivity.  The results on
$\swsqeffl$ are compared in Figure~\ref{fig:sin2teff}. The weighted
average of these six results, including small correlations, is:
\begin{eqnarray}
\swsqeffl & = & 0.23150\pm0.00016 \,.
\label{eq:sin2teff}
\end{eqnarray}
Note, however, that this average has a $\chi^2$ of 10.5 for 5 degrees
of freedom, corresponding to a probability of 6.2\%. The $\chi^2$ is
pushed up by the two most precise measurements of $\swsqeffl$, namely
those derived from the measurements of $\cAl$ by SLD, dominated by the
left-right asymmetry $\ALRz$, and of the forward-backward asymmetry
measured in $\bb$ production at LEP, $\Afbzb$, which differ by about
2.9 standard deviations. No experimental effect in either measurement
has been identified to explain this, thus the difference is presumably
either the effect of statistics or an unidentified systematics or a hint for new physics, as further discussed below.

Also shown in table 1 are the results on $m_W$ obtained at LEP-2 and at the Tevatron, and the new world average of the top mass.

For the analysis of electroweak data in the SM one starts from the
input parameters: as in any renormalisable theory masses and couplings
have to be specified from outside. One can trade one parameter for
another and this freedom is used to select the best measured ones as
input parameters. As a result, some of them, $\alpha$, $G_F$ and
$\MZ$, are very precisely known \cite{PDG}, some other ones, $m_{f_{light}}$,
$\Mt$ and $\alpha_s(m_Z)$ are far less well determined while $\MH$ is
largely unknown.  Note that the new combined CDF and D\O\ value for
$\Mt$~\cite{TEVEWWG-top}, as listed in Table~\ref{tab:msm:input}, is
higher than the previous average by nearly one standard deviation.

Among the light fermions, the quark masses are badly known, but
fortunately, for the calculation of radiative corrections, they can be
replaced by $\alpha(m_Z)$, the value of the QED running coupling at
the Z mass scale. The value of the hadronic contribution to the
running, $\dalhad$, reported in Table~\ref{tab:msm:input}, is obtained
through dispersion relations from the data on $\ee\to\rm{hadrons}$ at
low centre-of-mass energies~\cite{bib-BP01}. From the input parameters
one computes the radiative corrections to a sufficient precision to
match the experimental accuracy. Then one compares the theoretical
predictions and the data for the numerous observables which have been
measured, checks the consistency of the theory and derives constraints
on $\Mt$, $\alfmz$ and $\MH$.

The computed radiative corrections include the complete set of
one-loop diagrams, plus some selected large subsets of two-loop
diagrams and some sequences of resummed large terms of all orders
(large logarithms and Dyson resummations). In particular large
logarithms, e.g., terms of the form $(\alpha/\pi ~{\rm
ln}~(m_Z/m_{f_\ell}))^n$ where $f_{\ell}$ is a light fermion, are
resummed by well-known and consolidated techniques based on the
renormalisation group. For example, large logarithms dominate the
running of $\alpha$ from $m_e$, the electron mass, up to $\MZ$, which
is a $6\%$ effect, much larger than the few per-mille contributions of
purely weak loops.  Also, large logs from initial state radiation
dramatically distort the line shape of the Z resonance observed at
LEP-1 and SLC and must be accurately taken into account in the
measurement of the Z mass and total width.

Among the one loop EW radiative corrections a remarkable class of
contributions are those terms that increase quadratically with the top
mass.  The large sensitivity of radiative corrections to $m_t$ arises
from the existence of these terms. The quadratic dependence on $m_t$
(and possibly on other widely broken isospin multiplets from new
physics) arises because, in spontaneously broken gauge theories, heavy
loops do not decouple. On the contrary, in QED or QCD, the running of
$\alpha$ and $\alpha_s$ at a scale $Q$ is not affected by heavy quarks
with mass $M \gg Q$. According to an intuitive decoupling
theorem~\cite{AppCar}, diagrams with heavy virtual particles of mass
$M$ can be ignored for $Q \ll M$ provided that the couplings do not
grow with $M$ and that the theory with no heavy particles is still
renormalizable. In the spontaneously broken EW gauge theories both
requirements are violated. First, one important difference with
respect to unbroken gauge theories is in the longitudinal modes of
weak gauge bosons. These modes are generated by the Higgs mechanism,
and their couplings grow with masses (as is also the case for the
physical Higgs couplings). Second, the theory without the top quark is
no more renormalisable because the gauge symmetry is broken if the b
quark is left with no partner (while its measured couplings show that the weak
isospin is 1/2). Because of non decoupling precision tests of the
electroweak theory may be sensitive to new physics even if the new
particles are too heavy for their direct production.

While radiative corrections are quite sensitive to the top mass, they
are unfortunately much less dependent on the Higgs mass. If they were
sufficiently sensitive, by now we would precisely know the mass of the
Higgs. However, the dependence of one loop diagrams on $\MH$ is only
logarithmic: $\sim \GF\MW^2\log(\MH^2/\MW^2)$. Quadratic terms $\sim
\GF^2\MH^2$ only appear at two loops and are too small to be
important. The difference with the top case is that $\Mt^2-\Mb^2$ is a
direct breaking of the gauge symmetry that already affects the
relevant one loop diagrams, while the Higgs couplings to gauge bosons
are "custodial-SU(2)" symmetric in lowest order.

We now discuss fitting the data in the SM. One can think of different
types of fit, depending on which experimental results are included or
which answers one wants to obtain. For example, in
Table~\ref{tab:fit:result} we present in column~1 a fit of all Z pole
data plus $\MW$ and $\GW$ (this is interesting as it shows the value
of $m_t$ obtained indirectly from radiative corrections, to be
compared with the value of $m_t$ measured in production experiments),
in column~2 a fit of all Z pole data plus $\Mt$ (here it is $\MW$
which is indirectly determined), and, finally, in column~3 a fit of
all the data listed in Table~\ref{tab:msm:input} (which is the most
relevant fit for constraining $\MH$).  From the fit in column~1 of
Table~\ref{tab:fit:result} we see that the extracted value of $\Mt$ is
in perfect agreement with the direct measurement (see
Table~\ref{tab:msm:input}).  Similarly we see that the experimental
measurement of $\MW$ in Table~\ref{tab:msm:input} is larger by about
one standard deviation with respect to the value from the fit in
column~2.  We have seen that quantum corrections depend only
logarithmically on $\MH$.  In spite of this small sensitivity, the
measurements are precise enough that one still obtains a quantitative
indication of the mass range. From the fit in column~3 we obtain:
$\log_{10}{\MH(\GeV)}=2.05\pm 0.20$ (or $\MH=113^{+62}_{-42}~\GeV$).
This result on the Higgs mass is particularly remarkable. The value of
$\log_{10}{\MH(\GeV)}$ is right on top of the small window between
$\sim 2$ and $\sim 3$ which is allowed, on the one side, by the direct
search limit ($\MH\gappeq 114~\GeV$ from LEP-2~\cite{LEP2:MH-LIMIT}),
and, on the other side, by the theoretical upper limit on the Higgs
mass in the minimal SM, $\MH\lappeq 600-800~\GeV$ \cite{HR}.

\begin{table}
\begin{center}
\renewcommand{\arraystretch}{1.3}
\begin{tabular}{|l||c|c|c|}
\hline 
Fit       & 1 & 2 & 3 \\
\hline
\hline
Measurements      &$\MW$,~$\GW$          &$\Mt$            &$\Mt,~\MW,~\GW$\\
\hline
\hline
$\Mt~(\GeV)$      &$178.5^{+11.0}_{-8.5}$&$177.2\pm4.1$    &$178.1\pm3.9$\\
$\MH~(\GeV)$      &$117^{+162}_{-62}$    &$129^{+76}_{-50}$&$113^{+62}_{-42}$\\
$\log~[\MH(\GeV)]$&$2.07^{+0.38}_{-0.33}$&$2.11\pm0.21$    &$2.05\pm0.20$ \\
$\alpha_s(\MZ)$   &$0.1187\pm0.0027$     &$0.1190\pm0.0027$&$0.1186\pm0.0027$\\
\hline
$\chi^2/dof$      &$16.3/12$             &$15.0/11$        &$16.3/13$ \\
\hline
$\MW~(\MeV)$      &                      &$80386 \pm 23$   & \\
\hline
\end{tabular}
\caption[]{ Standard Model fits of electroweak data. All fits use the
Z pole results and $\dalhad$ as listed in Table~\ref{tab:msm:input},
also including constants such as the Fermi constant $\GF$. In
addition, the measurements listed in each column are included as
well. For fit~2, the expected W mass is also shown. For details on the
fit procedure see~\cite{LEPEWWG:2003}.}
\label{tab:fit:result}
\end{center}
\end{table}

\begin{figure}[t]
\begin{center}
\includegraphics[width=0.7\textwidth]{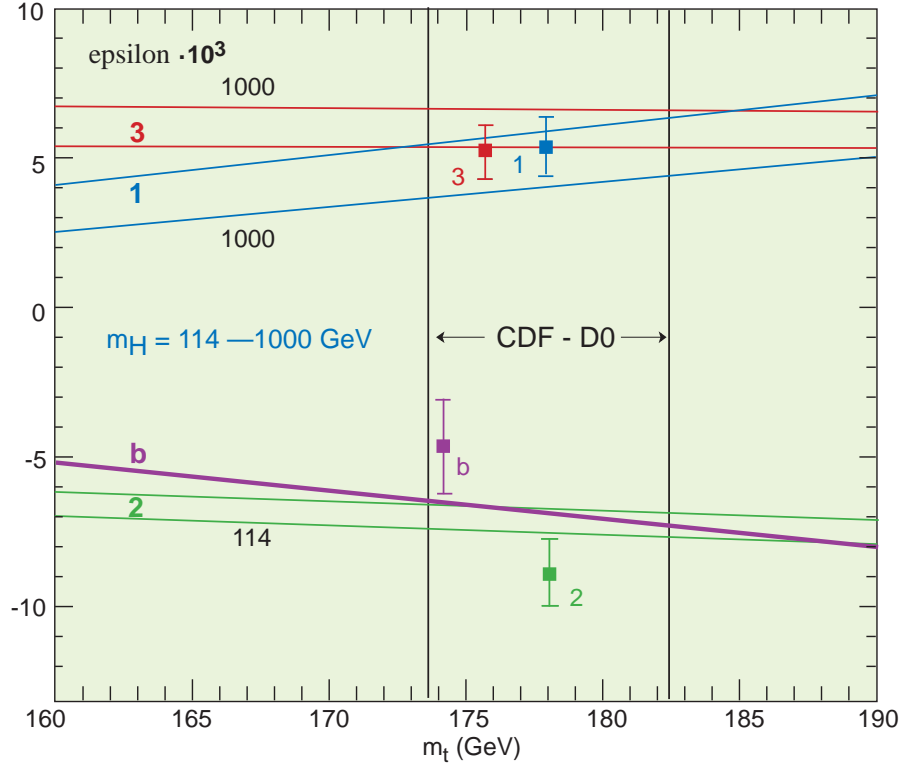}
\caption[]{ The epsilon variables: comparison of the data with the SM predictions.  The data should be horizontal bands but they are shown here near the central value of $m_t$.}
\label{fig:epsilon}
\end{center}
\end{figure}

A different way of looking at the data is to consider the epsilon parameters. As well known these parameters vanish in the limit of tree level SM plus pure QED or pure QCD corrections. So they are a measure of the weak quantum corrections. Their experimental values are given by \cite{LEPEW}:
\begin{eqnarray}
\epsilon_1 ~10^3& = & 5.4\pm1.0\\
\epsilon_2 ~10^3& = & -8.9\pm1.2 \\
\epsilon_3 ~10^3& = & 5.25\pm0.95 \\
\epsilon_b ~10^3& = & -4.7\pm1.6 
\label{eq:epsilons}
\end{eqnarray}
The experimental values are compared to the SM predictions as function of $m_t$ and $m_H$ in Figure~\ref{fig:epsilon}. We see that $\epsilon_3$ points to a light Higgs, that $\epsilon_b$ is a bit too large because of $A^b_{FB}$ and $\epsilon_2$ a bit too small because of $m_W$.

Thus the whole picture of a perturbative theory with a fundamental
Higgs is well supported by the data on radiative corrections. It is
important that there is a clear indication for a particularly light
Higgs: at $95\%$ c.l. $m_H\lappeq 237~\GeV$.  This is quite
encouraging for the ongoing search for the Higgs particle.  More
general, if the Higgs couplings are removed from the Lagrangian the
resulting theory is non renormalisable. A cutoff $\Lambda$ must be
introduced. In the quantum corrections $\log{m_H}$ is then replaced by
$\log{\Lambda}$ plus a constant. The precise determination of the
associated finite terms would be lost (that is, the value of the mass
in the denominator in the argument of the logarithm).  A heavy Higgs
would need some unfortunate conspiracy: the finite terms, different in
the new theory from those of the SM, should accidentally compensate
for the heavy Higgs in a few key parameters of the radiative
corrections (mainly $\epsilon_1$ and $\epsilon_3$, see, for example,
\cite{eps}).  Alternatively, additional new physics, for example in
the form of effective contact terms added to the minimal SM
lagrangian, should accidentally do the compensation, which again needs
some sort of conspiracy.

In Table~\ref{tab:msm:prediction} we collect the results on low energy
precision tests of the SM obtained from neutrino and antineutrino deep
inelastic scattering (NuTeV~\cite{bib-NuTeV-final}), parity violation
in Cs atoms (APV~\cite{QWCs:theo:2003}) and the recent measurement of
the parity-violating asymmetry in Moller scattering \cite{E158}.  The
experimental results are compared with the predictions from the fit in
column~3 of Table~\ref{tab:fit:result}.  We see the agreement is good
except for the NuTeV result that shows a deviation by three standard
deviations.  The NuTeV measurement is quoted as a measurement of
$\swsq=1-\MW^2/\MZ^2$ from the ratio of neutral to charged current
deep inelastic cross-sections from $\nu_{\mu}$ and $\bar{\nu}_{\mu}$
using the Fermilab beams. There is growing evidence that the NuTeV
anomaly could simply arise from an underestimation of the theoretical
uncertainty in the QCD analysis needed to extract $\swsq$.  In fact,
the lowest order QCD parton formalism on which the analysis has been
based is too crude to match the experimental accuracy.  In particular
a small asymmetry in the momentum carried by the strange and
antistrange quarks, $s-\bar s$, could have a large effect
\cite{DFGRS}. A tiny violation of isospin symmetry in parton
distributions, too small to be seen elsewhere, can similarly be of
some importance. In conclusion we believe the discrepancy has more to
teach about the QCD parton densities than about the electroweak
theory.

\begin{table}
\begin{center}
  \renewcommand{\arraystretch}{1.30}
\begin{tabular}{|ll||r||r|}
\hline
&Observable& {Measurement}  & {SM fit}  \\
\hline
\hline
&$\swsq$        ($\nu$N~\cite{bib-NuTeV-final})
                & $0.2277\pm0.0016\pz$  & 0.2226$\pz$ \\
\hline
&$\QWCs$ (APV~\cite{QWCs:theo:2003})
                & $-72.83\pm0.49\pzz\pz$& $-72.91\pzz\pz$ \\
\hline
&$\swsqeffl$   ($\mathrm{e^-e^-}$~\cite{E158})
                & $0.2296\pm0.0023\pz$  & 0.2314$\pz$ \\
\hline
%-----------------------------------------------------------------------
\end{tabular}\end{center}
\caption[Overview of results]{ Summary of other electroweak precision
measurements, namely the measurements of the on-shell electroweak
mixing angle in neutrino-nucleon scattering, the weak charge of cesium
measured in an atomic parity violation experiment, and the effective
weak mixing angle measured in Moller scattering, all performed in
processes at low $Q^2$.  The SM predictions are derived from fit 3 of
Table~\ref{tab:fit:result}.  Good agreement of the prediction with the
measurement is found except for $\nu$N. }
\label{tab:msm:prediction}
\end{table}

When confronted with these results, on the whole the SM performs
rather well, so that it is fair to say that no clear indication for
new physics emerges from the data. However, as already mentioned, one
problem is that the two most precise measurements of $\swsqeffl$ from
$\ALR$ and $\Afbzb$ differ nearly three standard deviations.  In
general, there appears to be a discrepancy between $\swsqeffl$
measured from leptonic asymmetries ($(\sin^2\theta_{\rm eff})_l$) and
from hadronic asymmetries ($(\sin^2\theta_{\rm eff})_h$), see also
Figure~\ref{fig:sin2teff}. In fact, the result from $\ALR$ is in good
agreement with the leptonic asymmetries measured at LEP, while all
hadronic asymmetries, though their errors are large, are better
compatible with the result of $\Afbzb$.

\begin{figure}[t]
\begin{center}
\includegraphics[width=0.7\textwidth]{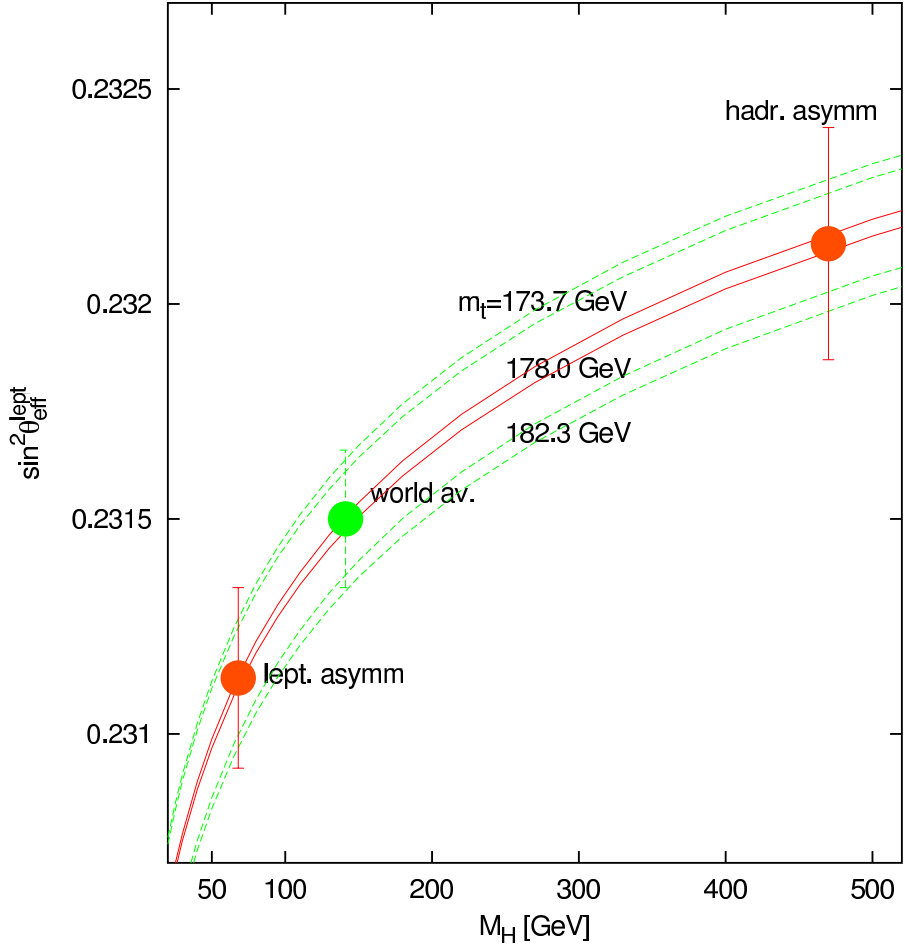}
\caption[]{ The data for $\sin^2\theta_{\rm eff}^{\rm lept}$ are
plotted vs $m_H$. For presentation purposes the measured points are
shown each at the $m_H$ value that would ideally correspond to it
given the central value of $m_t$ (updated from \cite{P-Gambino}).}
\label{fig:higgs}
\end{center}
\end{figure} 

The situation is shown in Figure~\ref{fig:higgs}~\cite{P-Gambino}.
The values of $(\sin^2\theta_{\rm eff})_l$, $(\sin^2\theta_{\rm
eff})_h$ and their formal combination are shown each at the $\MH$
value that would correspond to it given the central value of $\Mt$.
Of course, the value for $\MH$ indicated by each $\swsqeffl$ has an
horizontal ambiguity determined by the measurement error and the width
of the $\pm1\sigma$ band for $\Mt$.  Even taking this spread into
account it is clear that the implications on $\MH$ are sizably
different.  One might imagine that some new physics effect could be
hidden in the $\mathrm{Z b \bar b}$ vertex.  Like for the top quark
mass there could be other non decoupling effects from new heavy states
or a mixing of the b quark with some other heavy quark.  However, it
is well known that this discrepancy is not easily explained in terms
of some new physics effect in the $\mathrm{Z b \bar b}$ vertex. In
fact, $\Afbzb$ is the product of lepton- and b-asymmetry factors:
$\Afbzb=(3/4)\cAe\cAb$.  The sensitivity of $\Afbzb$ to $\cAb$ is
limited, because the $\cAe$ factor is small, so that a rather large
change of the b-quark couplings with respect to the SM is needed in
order to reproduce the measured discrepancy (precisely a $\sim 30\%$
change in the right-handed coupling, an effect too large to be a loop
effect but which could be produced at the tree level, e.g., by mixing
of the b quark with a new heavy vectorlike quark \cite{CTW}).  But
then this effect should normally also appear in the direct measurement
of $\cAb$ performed at SLD using the left-right polarized b asymmetry,
even within the moderate precision of this result, and it should also
be manifest in the accurate measurement of $\Rb \propto
g_{\mathrm{Rb}}^2+g_{\mathrm{Lb}}^2$.  The measurements of neither
$\cAb$ nor $\Rb$ confirm the need of a new effect. Even introducing an
ad hoc mixing the overall fit is not terribly good, but we cannot
exclude this possibility completely.  Alternatively, the observed
discrepancy could be due to a large statistical fluctuation or an
unknown experimental problem. The ambiguity in the measured value of
$\swsqeffl$ could thus be larger than the nominal error, reported in
Equation~\ref{eq:sin2teff}, obtained from averaging all the existing
determinations.

\begin{figure}[t]
\begin{center}
\includegraphics[width=0.8\textwidth]{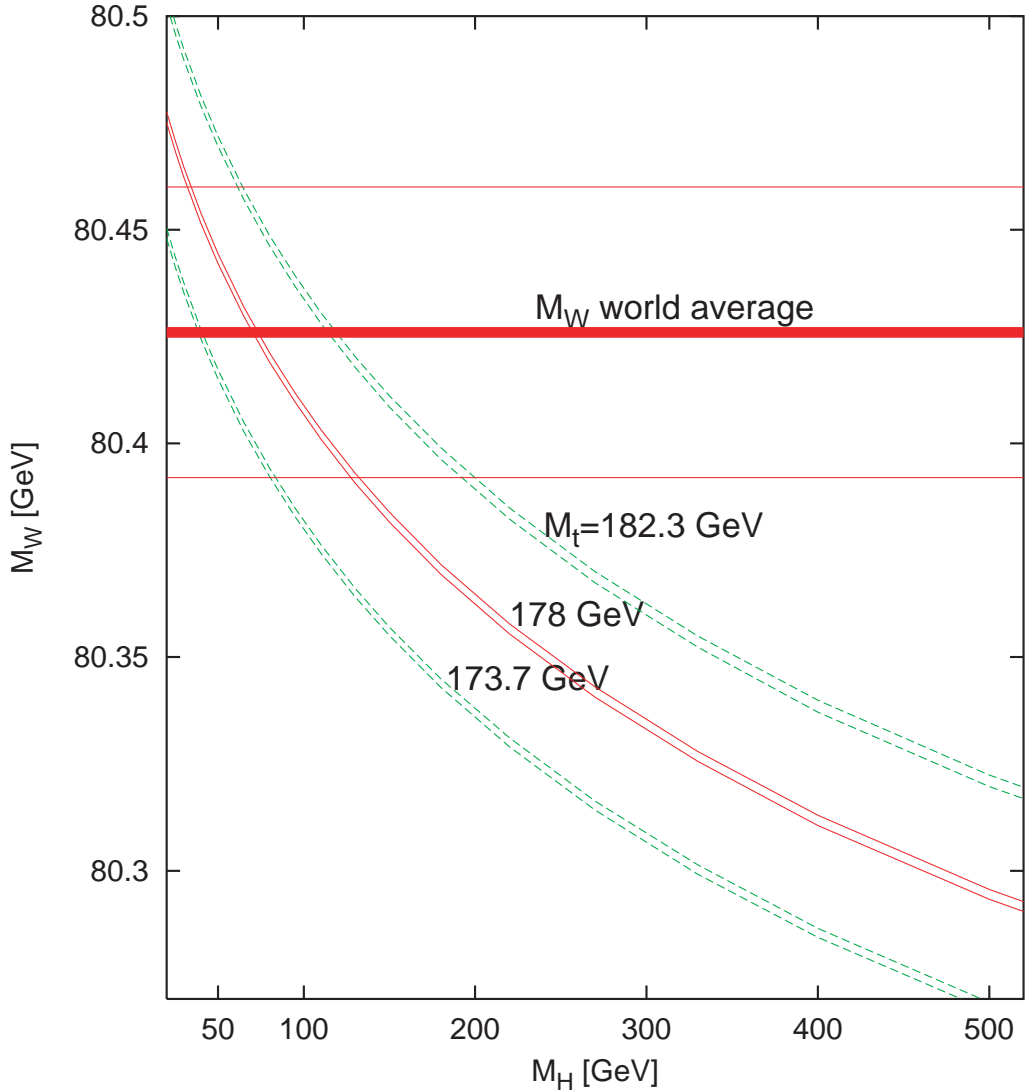}
\caption[]{The world average for $m_W$ is compared with the SM prediction as a function of $m_H$ (updated from \cite{P-Gambino}). }
\label{fig:GambinoW04}
\end{center}
\end{figure}

We have already observed that the experimental value of $\MW$ (with
good agreement between LEP and the Tevatron) is a bit high compared to
the SM prediction (see Figure~\ref{fig:GambinoW04}). The value of $\MH$
indicated by $\MW$ is on the low side, just in the same interval as
for $\sin^2\theta_{\rm eff}^{\rm lept}$ measured from leptonic
asymmetries.  It is interesting that the new value of $\Mt$
considerably relaxes the previous tension between the experimental
values of $\MW$ and $\sin^2\theta_{\rm eff}^{\rm lept}$ measured from
leptonic asymmetries on one side and the lower limit on $\MH$ from
direct searches on the other side~\cite{cha,ACGGR}.  This is also
apparent from Figure~\ref{fig:GambinoW04}.

The main lesson of precision tests of the standard electroweak theory
can be summarised as follows. The couplings of quark and leptons to
the weak gauge bosons W$^{\pm}$ and Z are indeed precisely those
prescribed by the gauge symmetry. The accuracy of a few per-mille for
these tests implies that, not only the tree level, but also the
structure of quantum corrections has been verified. To a lesser
accuracy the triple gauge vertices $\gamma\WW$ and Z$\WW$ have also
been found in agreement with the specific prediction of the
$SU(2)\bigotimes U(1)$ gauge theory. This means that it has been
verified that the gauge symmetry is unbroken in the vertices of the
theory: the currents are indeed conserved. Yet there is obvious
evidence that the symmetry is otherwise badly broken in the
masses. Thus the currents are conserved but the spectrum of particle
states is not at all symmetric. This is a clear signal of spontaneous
symmetry breaking. The practical implementation of spontaneous
symmetry breaking in a gauge theory is via the Higgs mechanism. The
Higgs sector of the SM is still very much untested. What has been
tested is the relation $\MW^2=\MZ^2\cwsq$, modified by computable
radiative corrections. This relation means that the effective Higgs
(be it fundamental or composite) is indeed a weak isospin doublet.
The Higgs particle has not been found but in the SM its mass can well
be larger than the present direct lower limit $m_H\gappeq114~\GeV$
obtained from direct searches at LEP-2.  The radiative corrections
computed in the SM when compared to the data on precision electroweak
tests lead to a clear indication for a light Higgs, not too far from
the present lower bound. No signal of new physics has been
found. However, to make a light Higgs natural in presence of quantum
fluctuations new physics should not be too far. This is encouraging
for the LHC that should experimentally clarify the problem of the
electroweak symmetry breaking sector and search for physics beyond the
SM.

\section{Outlook on Avenues beyond the Standard Model}

Given the success of the SM why are we not satisfied with that theory? Why not just find the Higgs particle,
 for completeness, and declare that particle physics is closed? The reason is that there are
both conceptual problems and phenomenological indications for physics beyond the SM. On the conceptual side the most
obvious problems are that quantum gravity is not included in the SM and the related hierarchy problem. Among the main
phenomenological hints for new physics we can list coupling unification, dark matter, neutrino masses, 
baryogenesis and the cosmological vacuum energy.

The computed evolution with energy
of the effective SM gauge couplings clearly points towards the unification of the electro-weak and strong forces (Grand
Unified Theories: GUT's) at scales of energy
$M_{GUT}\sim  10^{15}-10^{16}~ GeV$ which are close to the scale of quantum gravity, $M_{Pl}\sim 10^{19}~ GeV$.  One is led to
imagine  a unified theory of all interactions also including gravity (at present superstrings provide the best attempt at such
a theory). Thus GUT's and the realm of quantum gravity set a very distant energy horizon that modern particle theory cannot
ignore. Can the SM without new physics be valid up to such large energies? This appears unlikely because the structure of the
SM could not naturally explain the relative smallness of the weak scale of mass, set by the Higgs mechanism at $\mu\sim
1/\sqrt{G_F}\sim  250~ GeV$  with $G_F$ being the Fermi coupling constant. This so-called hierarchy problem is related to
the
presence of fundamental scalar fields in the theory with quadratic mass divergences and no protective extra symmetry at
$\mu=0$. For fermion masses, first, the divergences are logarithmic and, second, they are forbidden by the $SU(2)\bigotimes
U(1)$ gauge symmetry plus the fact that at
$m=0$ an additional symmetry, i.e. chiral  symmetry, is restored. Here, when talking of divergences, we are not
worried of actual infinities. The theory is renormalisable and finite once the dependence on the cut off is
absorbed in a redefinition of masses and couplings. Rather the hierarchy problem is one of naturalness. We should see the
cut off as a parameterization of our ignorance on the new physics that will modify the theory at large energy
scales. Then it is relevant to look at the dependence of physical quantities on the cut off and to demand that no
unexplained enormously accurate cancellations arise. 

The hierarchy problem can be put in very practical terms: loop corrections to the higgs mass squared are
quadratic in $\Lambda$. The most pressing problem is from the top loop.
 With $m_h^2=m^2_{bare}+\delta m_h^2$ the top loop gives 
 \begin{eqnarray}
\delta m_{h|top}^2\sim \frac{3G_F}{\sqrt{2} \pi^2} m_t^2 \Lambda^2\sim(0.3\Lambda)^2 \label{top}
\end{eqnarray}

If we demand that the correction does not exceed the light Higgs mass indicated by the precision tests, $\Lambda$ must be
close, $\Lambda\sim o(1~TeV)$. Similar constraints arise from the quadratic $\Lambda$ dependence of loops with gauge bosons and
scalars, which, however, lead to less pressing bounds. So the hierarchy problem demands new physics to be very close (in
particular the mechanism that quenches the top loop). Actually, this new physics must be rather special, because it must be
very close, yet its effects are not clearly visible (the "LEP Paradox" \cite{BS}). Examples of proposed classes of solutions
for the hierarchy problem are:

¥ $\bf{Supersymmetry.}$ In the limit of exact boson-fermion symmetry the quadratic divergences of bosons cancel so that
only log divergences remain. However, exact SUSY is clearly unrealistic. For approximate SUSY (with soft breaking terms),
which is the basis for all practical models, $\Lambda$ is replaced by the splitting of SUSY multiplets, $\Lambda\sim
m_{SUSY}-m_{ord}$. In particular, the top loop is quenched by partial cancellation with s-top exchange. 

¥ $\bf{Technicolor.}$ The Higgs system is a condensate of new fermions. There are no fundamental scalar Higgs sector, hence no
quadratic devergences associated to the $\mu^2$ mass in the scalar potential. This mechanism needs a very strong binding force,
$\Lambda_{TC}\sim 10^3~\Lambda_{QCD}$. It is  difficult to arrange that such nearby strong force is not showing up in
precision tests. Hence this class of models has been disfavoured by LEP, although some special class of models have been devised aposteriori, like walking TC, top-color assisted TC etc (for
recent reviews, see, for example, \cite{L-C}).

  $\bf{Large~compactified~extra~dimensions.}$ The idea is that $M_{PL}$ appears very large, that is gravity seems very weak
because we are fooled by hidden extra dimensions so that the real gravity scale is reduced down to
$o(1~TeV)$. This possibility is very exciting in itself and it is really remarkable that it is compatible with experiment.

¥ $\bf{"Little~Higgs" models.}$ In these models extra symmetries allow $m_h\not= 0$ only at two-loop level, so that $\Lambda$
can be as large as
$o(10~TeV)$ with the Higgs within present bounds (the top loop is quenched by exchange of heavy vectorlike new charge-2/3 quarks).

We now briefly comment in turn on these possibilities.

SUSY models are the most developed and most widely accepted. Many theorists consider SUSY as established at the Planck
scale $M_{Pl}$. So why not to use it also at low energy to fix the hierarchy problem, if at all possible? It is interesting
that viable models exist. The necessary SUSY breaking can be introduced through soft
terms that do not spoil the good convergence properties of the theory. Precisely those terms arise from
supergravity when it is spontaneoulsly broken in a hidden sector. This is the case of the MSSM \cite{Martin}. Of course, minimality is only a simplicity assumption that could possibly be relaxed. The MSSM 
is a completely specified,
consistent and computable theory which is compatible with all precision electroweak tests. In this
most traditional approach SUSY is broken in a hidden sector and the scale of SUSY breaking is very
large of order
$\Lambda\sim\sqrt{G^{-1/2}_F M_{Pl}}$. But since the hidden sector only communicates with the visible sector
through gravitational interactions the splitting of the SUSY multiplets is much smaller, in the TeV
energy domain, and the Goldstino is practically decoupled. 
But alternative mechanisms of SUSY breaking are also being considered. In one alternative scenario \cite{gau} the (not so
much) hidden sector is connected to the visible one by ordinary gauge interactions. As these are much
stronger than the gravitational interactions, $\Lambda$ can be much smaller, as low as 10-100
TeV. It follows that the Goldstino is very light in these models (with mass of order or below 1 eV
typically) and is the lightest, stable SUSY particle, but its couplings are observably large. The radiative
decay of the lightest neutralino into the Goldstino leads to detectable photons. The signature of photons comes
out naturally in this SUSY breaking pattern: with respect to the MSSM, in the gauge mediated model there are typically
more photons and less missing energy. The main appeal of gauge mediated models is a better protection against
flavour changing neutral currents but naturality problems tend to increase. As another possibility it has been
pointed out that there are pure gravity contributions to soft masses that arise from gravity theory anomalies\cite{ano}. In the assumption that these terms are dominant the associated spectrum and phenomenology have been
studied. In this case gaugino masses are proportional to gauge coupling beta functions, so that the gluino is much heavier
than the electroweak gauginos, and the wino is most often the lightest SUSY particle. 

What is really unique to SUSY with respect to all other extensions of the SM listed above is that the MSSM or
similar models are well defined and computable up to $M_{Pl}$ and, moreover, are not only compatible but actually
quantitatively supported by coupling unification and GUT's. At present the most direct
phenomenological evidence in favour of supersymmetry is obtained from the unification of couplings in GUTs.
Precise LEP data on $\alpha_s(m_Z)$ and $\sin^2{\theta_W}$ show that
standard one-scale GUTs fail in predicting $\sin^2{\theta_W}$ given
$\alpha_s(m_Z)$ (and $\alpha(m_Z)$) while SUSY GUTs are in agreement with the present, very precise,
experimental results. If one starts from the known values of
$\sin^2{\theta_W}$ and $\alpha(m_Z)$, one finds \cite{LP} for $\alpha_s(m_Z)$ the results:
$\alpha_s(m_Z) = 0.073\pm 0.002$ for Standard GUTs and $\alpha_s(m_Z) = 0.129\pm0.010$ for SUSY~ GUTs
to be compared with the world average experimental value $\alpha_s(m_Z) =0.119\pm0.003$. Another great asset of SUSY GUT's
is that proton decay is much slowed down with respect to the non SUSY case. First, the unification mass $M_{GUT}\sim~\rm{few}~
10^{16}~GeV$, in typical SUSY GUT's, is about 20-30 times larger than for ordinary GUT's. This makes p decay via gauge
boson exchange negligible and the main decay amplitude arises from dim-5 operators with higgsino exchange, leading to a
rate close but still compatible with existing bounds (see, for example,\cite{AFM}). It is also important that SUSY provides an excellent dark matter candidate, the neutralino.  We finally recall that the range of neutrino masses as indicated by oscillation experiments, when interpreted in the see-saw mechanism, point to $M_{GUT}$ and give additional support to GUTs \cite{alfe}.

In spite of all these virtues it is true that the lack of SUSY signals at LEP and the lower limit on $m_H$ pose problems
for the MSSM. The lightest Higgs particle is predicted in the MSSM to be below $m_h\lappeq~135~GeV$ (the recent increase of $m_t$ helps in this respect). The limit on the SM
Higgs
$m_H\gappeq~114~GeV$ considerably restricts the available parameter space of the MSSM requiring relatively large $\tan\beta$
($\tan\beta\gappeq~2-3$: at tree level $m^2_h=m^2_Z\cos^2{2\beta}$) and rather heavy s-top (the loop corrections
increase with $\log{\tilde{m_t^2}}$). Stringent naturality constraints also follow from imposing that the electroweak
symmetry breaking occurs at the right place: in SUSY models the breaking is induced by the running of the $H_u$ mass
starting from a common scalar mass $m_0$ at $M_{GUT}$. The squared Z mass $m_Z^2$ can be expressed as a linear
combination of the SUSY parameters $m_0^2$, $m_{1/2}^2$, $A^2_t$, $\mu^2$,... with known coefficients. Barring
cancellations that need fine tuning, the SUSY parameters, hence the SUSY s-partners cannot be too heavy. The LEP limits,
in particular the chargino lower bound $m_{\chi+}\gappeq~100~GeV$, are sufficient to eliminate an important region of the
parameter space, depending on the amount of allowed fine tuning. For example, models based on gaugino universality at the
GUT scale are discarded unless a fine tuning by at least a factor of ~20 is not allowed. Without gaugino
universality \cite{kane} the strongest limit remains on the gluino mass: $m_Z^2\sim 0.7~m_{gluino}^2+\dots$ which is still
compatible with the present limit $m_{gluino}\gappeq~200~GeV$.

The non discovery of SUSY at LEP has given further impulse to the quest for new ideas on physics beyond the SM. Large extra
dimensions \cite{Jo} and "little Higgs" \cite{schm} models are the most interesting new directions in model building. Large
extra dimension models propose to solve the hierarchy problem by bringing gravity down from $M_{Pl}$ to $m\sim~o(1~TeV)$ where
$m$ is the string scale. Inspired by string theory one assumes that some compactified extra dimensions are sufficiently large
and that the SM fields are confined to a 4-dimensional brane immersed in a d-dimensional bulk while gravity, which feels the
whole geometry, propagates in the bulk. We know that the Planck mass is large because gravity is weak: in fact $G_N\sim
1/M_{Pl}^2$, where
$G_N$ is Newton constant. The idea is that gravity appears so weak because a lot of lines of force escape in extra
dimensions. Assume you have $n=d-4$ extra dimensions with compactification radius $R$. For large distances, $r>>R$, the
ordinary Newton law applies for gravity: in natural units $F\sim G_N/r^2\sim 1/(M_{Pl}^2r^2)$. At short distances,
$r\lappeq R$, the flow of lines of force in extra dimensions modifies Gauss law and $F^{-1}\sim m^2(mr)^{d-4}r^2$. By
matching the two formulas at $r=R$ one obtains $(M_{Pl}/m)^2=(Rm)^{d-4}$. For $m\sim~1~TeV$ and $n=d-4$ one finds that
$n=1$ is excluded ($R\sim 10^{15} cm$), for $n=2~R$  is at the edge of present bounds $R\sim~1~ mm$, while for $n=4,6$,
$R\sim~10^{-9}, 10^{-12}~cm$.  In all these models a generic feature is the occurrence of Kaluza-Klein (KK) modes.
Compactified dimensions with periodic boundary conditions, as for quantization in a box, imply a discrete spectrum with
momentum $p=n/R$ and mass squared $m^2=n^2/R^2$. There are many versions of these models. The SM brane can itself have a
thickness $r$ with $r<\sim~10^{-17}~cm$ or $1/r>\sim~1~TeV$, because we know that quarks and leptons are pointlike down to
these distances, while for gravity there is no experimental counter-evidence down to $R<\sim~0.1~mm$ or
$1/R>\sim~10^{-3}~eV$. In case of a thickness for the SM brane there would be KK recurrences for SM fields, like $W_n$,
$Z_n$ and so on in the $TeV$ region and above. There are models with factorized metric ($ds^2=\eta_{\mu
\nu}dx^{\mu}dx^{\nu}+h_{ij}(y)dy^idy^j$, where y (i,j) denotes the extra dimension coordinates (and indices), or models
with warped metric ($ds^2=e-{2kR|\phi|} \eta_{\mu \nu}dx^{\mu}dx^{\nu}-R^2\phi^2$ \cite{RS}.
In any case there are the towers of KK recurrences of the graviton. They
are gravitationally coupled but there are a lot of them that sizably couple, so that the net result is a modification
of cross-sections and the presence of missing energy. 

Large extra dimensions provide a very exciting scenario \cite{FeAa}. Already it is remarkable that this possibility is
compatible with experiment. However, there are a number of criticisms that can be brought up. First, the hierarchy problem is
more translated in new terms rather than solved. In fact the basic relation $Rm=(M_{Pl}/m)^{2/n}$ shows that $Rm$, which one
would apriori expect to be $0(1)$, is instead ad hoc related to the large ratio $M_{Pl}/m$.  In this respect  the Randall-Sundrum variety is more appealing because the hierarchy suppression $m_W/M_{Pl}$ could arise from the warping factor $e^{-2kR|\phi|}$, with not too large values of $kR$. Also it is
not clear how extra dimensions can by themselves solve the LEP paradox (the large top loop corrections should be
controlled by the opening of the new dimensions and the onset of gravity): since
$m_H$ is light
$\Lambda\sim 1/R$ must be relatively close. But precision tests put very strong limits on $\Lambda$ In fact in typical
models of this class there is no mechanism to sufficiently quench the corrections. No simple, realistic model has
yet emerged as a benchmark. But it is attractive to imagine that large extra dimensions could be a part of the truth,
perhaps coupled with some additional symmetry or even SUSY.

In the extra dimension general context an interesting direction of development is the study of symmetry breaking by orbifolding and/or boundary conditions. These are models where a larger gauge symmetry (with or without SUSY) holds in the bulk. The symmetry is reduced in the 4 dimensional brane, where the physics that we observe is located, as an effect of symmetry breaking induced geometrically by suitable boundary conditions. There are models where SUSY, valid in $n>4$ dimensions is broken by boundary conditions \cite{ant}, in particular the model of ref.\cite{bar}, where the mass of the Higgs is computable and can be extimated with good accuracy. Then there are "Higgsless  models" where it is the SM electroweak gauge symmetry which is broken at the boundaries \cite{Hless}.  Or models where the Higgs is the 5th component of a gauge boson of an extended  symmetry valid in $n>4$ \cite{hoso}. In general all these alternative models for the Higgs mechanism face severe problems and constraints from electroweak precision tests \cite{BPR}. At the GUT scale, symmetry breaking by orbifolding can be applied to obtain a reformulation of SUSY GUT's where many problematic features of ordinary GUT's (e.g. a baroque Higgs sector, the doublet-triplet splitting problem, fast proton decay etc) are improved \cite{Kaw}, \cite{FeAa}.

In "little Higgs" models the symmetry of the SM is extended to a suitable global group G that also contains some
gauge enlargement of $SU(2)\bigotimes U(1)$, for example $G\supset [SU(2)\bigotimes U(1)]^2\supset SU(2)\bigotimes
U(1)$. The Higgs particle is a pseudo-Goldstone boson of G that only takes mass at 2-loop level, because two distinct
symmetries must be simultaneously broken for it to take mass,  which requires the action of two different couplings in
the same diagram. Then in the relation
between
$\delta m_h^2$ and
$\Lambda^2$ there is an additional coupling and an additional loop factor that allow for a bigger separation between the Higgs
mass and the cut-off. Typically, in these models one has one or more Higgs doublets at $m_h\sim~0.2~TeV$, and a cut-off at
$\Lambda\sim~10~TeV$. The top loop quadratic cut-off dependence is partially canceled, in a natural way guaranteed by the
symmetries of the model, by a new coloured, charge-2/3, vectorial quark $\chi$ of mass around $1~TeV$ (a fermion not a scalar
like the s-top of SUSY models). Certainly these models involve a remarkable level of group theoretic virtuosity. However, in
the simplest versions one is faced with problems with precision tests of the SM \cite{prob}. Even with
vectorlike new fermions large corrections to the epsilon parameters arise from exchanges of the new gauge bosons
$W'$ and $Z'$ (due to lack of custodial $SU(2)$ symmetry). In order to comply with these constraints the cut-off must be
pushed towards large energy and the amount of fine tuning needed to keep the Higgs light is still quite large.
Probably these bad features can be fixed by some suitable complication of the model (see for example, \cite{Ch}). But, in my opinion, the real limit of
this approach is that it only offers a postponement of the main problem by a few TeV, paid by a complete loss of
predictivity at higher energies. In particular all connections to GUT's are lost. 

Finally, we stress the importance of  the cosmological constant or vacuum energy problem \cite{tu}. The exciting recent results
on cosmological parameters, culminating with the precise WMAP measurements \cite{WMAP}, have shown that vacuum energy accounts
for about 2/3 of the critical density: $\Omega_{\Lambda}\sim 0.65$, Translated into familiar units this means for the energy
density $\rho_{\Lambda}\sim (2~10^{-3}~eV)^4$ or $(0.1~mm)^{-4}$. It is really interesting (and not at all understood)
that $\rho_{\Lambda}^{1/4}\sim \Lambda_{EW}^2/M_{Pl}$ (close to the range of neutrino masses). It is well known that in
field theory we expect $\rho_{\Lambda}\sim \Lambda_{cutoff}^4$. If the cut off is set at $M_{Pl}$ or even at $0(1~TeV)$
there would an enormous mismatch. In exact SUSY $\rho_{\Lambda}=0$, but SUSY is broken and in presence of breaking 
$\rho_{\Lambda}^{1/4}$ is in general not smaller than the typical SUSY multiplet splitting. Another closely related
problem is "why now?": the time evolution of the matter or radiation density is quite rapid, while the density for a
cosmological constant term would be flat. If so, them how comes that precisely now the two density sources are
comparable? This suggests that the vacuum energy is not a cosmological constant term, buth rather the vacuum expectation
value of some field (quintessence) and that the "why now?" problem is solved by some dynamical mechanism. 

Clearly the cosmological constant problem poses a big question mark on the relevance of naturalness as a relevant criterion also for the hierarchy problem: how we can trust that we need new physics close to the weak scale out of naturalness if we have no idea on the solution of the cosmological constant huge naturalness problem? The common answer is that the hierarchy problem is formulated within a well defined field theory context while the cosmological constant problem makes only sense within a theory of quantum gravity, that there could be modification of gravity at the sub-eV scale, that the vacuum energy could flow in extra dimensions or in different Universes and so on. At the other extreme is the possibility that naturalness is misleading. Weinberg \cite{We} has pointed out that the observed order of magnitude of $\Lambda$ can be successfully reproduced as the one necessary to allow galaxy formation in the Universe. In a scenario where new Universes are continuously produced we might be living in a very special one (largely fine-tuned) but the only one to allow the development of an observer. One might then argue that the same could in principle be true also for the Higgs sector. Recently it was suggested \cite{AHD} to abandon the no-fine-tuning assumption for the electro-weak theory, but require correct coupling unification, presence of dark matter with weak couplings and a single scale of evolution from the EW to the GUT scale. A "split SUSY" model arises as a solution with a fine-tuned light Higgs and all SUSY particles heavy except for gauginos, higgsinos and neutralinos, protected by chiral symmetry. Or we can have a two-scale non-SUSY GUT with axions as dark matter. In conclusion, it is clear that naturalness can be a good heuristic principle but you cannot prove its necessity.

\section{Summary and Conclusion}

Supersymmetry remains the standard way beyond the SM. What is unique to SUSY, beyond leading to a set of consistent and
completely formulated models, as, for example, the MSSM, is that this theory can potentially work up to the GUT energy scale.
In this respect it is the most ambitious model because it describes a computable framework that could be valid all the way
up to the vicinity of the Planck mass. The SUSY models are perfectly compatible with GUT's and are actually quantitatively
supported by coupling unification and also by what we have recently learned on neutrino masses. All other main ideas for going
beyond the SM do not share this synthesis with GUT's. The SUSY way is testable, for example at the LHC, and the issue
of its validity will be decided by experiment. It is true that we could have expected the first signals of SUSY already at
LEP, based on naturality arguments applied to the most minimal models (for example, those with gaugino universality at
asymptotic scales). The absence of signals has stimulated the development of new ideas like those of large extra dimensions
and "little Higgs" models. These ideas are very interesting and provide an important referfence for the preparation of LHC
experiments. Models along these new ideas are not so completely formulated and studied as for SUSY and no well defined and
realistic baseline has sofar emerged. But it is well possible that they might represent at least a part of the truth and it
is very important to continue the exploration of new ways beyond the SM.

I would like to express my gratitude to the Organisers of the ICFA '03 School for their invitation and their magnificent hospitality in Itacuruca. In particular I would like to thank Bernard Marechal.

\end{document}